\documentclass[pra]{revtex4}
 \usepackage{amssymb} \usepackage{graphicx}
\usepackage[utf8]{inputenc}
\usepackage[english]{babel} 
\usepackage{amsthm}
 \usepackage{amsmath}

\begin{document}
 \title{Algebra structure of conformal Killing-Yano forms in geometries with skew-symmetric torsion}

\author{\"{U}mit Ertem}
\altaffiliation{Corresponding author: \texttt{umitertem@ankara.edu.tr}}
\affiliation{Department of Software Engineering, Ankara University,\\
Faculty of Engineering, 06830 G\"olba\c{s}{\i}, Ankara, Turkey}

\author{\"{O}zg\"{u}r Kelek\c{c}i}
\email{okelekci@thk.edu.tr}
\affiliation{Department of Basic Sciences, Faculty of Engineering,\\
 University of Turkish Aeronautical Association, Ankara, Turkey}

\author{\"{O}zg\"{u}r A\c{c}{\i}k}
\email{ozacik@science.ankara.edu.tr}
\affiliation{Department of Physics, Ankara University,\\
 Faculty of Sciences, 06100, Tandogan-Ankara, Turkey}

\begin{abstract}

We consider conformal Killing-Yano forms corresponding to the antisymmetric generalizations of conformal Killing vectors to higher degree forms in the presence of skew-symmetric torsion. Integrability conditions for torsionful conformal Killing-Yano forms are found and a graded Lie bracket for conformal Killing-Yano forms to constitute a graded Lie algebra structure is proposed. It is found that a graded Lie algebra structure for a special subset of torsionful conformal Killing-Yano forms can be constructed for a closed and parallel skew-symmetric torsion on constant curvature and Einstein manifolds. Similar structure for generalized hidden symmetries defined from generalized connection in generalized geometry is also constructed.

\end{abstract}

\maketitle

\section{Introduction}

Isometries of a manifold are generated by Killing vector fields and they constitute a Lie algebra structure under the Lie bracket of vector fields. Similarly, the vector fields that preserve the metric up to conformal transformations correspond to conformal Killing vector fields and they also satisfy a Lie algebra structure. Antisymmetric generalizations of Killing and conformal Killing vector fields to higher degree differential forms are called Killing-Yano (KY) and conformal Killing-Yano (CKY) forms, respectively \cite{Yano, TachibanaKashiwada, Semmelmann}. They are called hidden symmetries of the manifold in general and used in various areas of mathematical physics such as in the construction of the first integrals of the geodesic motion \cite{HughstonPenroseSommersWalker, KrtousKubiznakPageFrolov}, symmetry operators of the massless and massive Dirac equation \cite{BennKress, AcikErtemOnderVercin1, BennCharlton, CarigliaKrtousKubiznak, Cariglia}, gravitational currents related to $p$-branes \cite{KastorTraschen, AcikErtemOnderVercin2, ErtemAcik} and extension of symmetry and conformal superalgebras related to geometric Killing spinors and twistor spinors \cite{Ertem, Ertem2, AcikErtem}. The question for the extension of the Lie algebra structure of Killing and conformal Killing vector fields to the higher degree generalizations has recently been studied in the literature \cite{AcikErtem, KastorRayTraschen, Ertem3, CarigliaKrtousKubiznak2}. KY forms constitute a graded Lie algebra structure on constant curvature manifolds under Schouten-Nijenhuis (SN) bracket. However, for CKY forms to satisfy a graded Lie algebra structure one has to define a new bracket constructed from the modification of the SN bracket and the graded Lie algebra structure is satisfied for all CKY forms on constant curvature manifolds and for normal CKY forms on Einstein manifolds \cite{Ertem3}. Normal CKY forms correspond to a subset of CKY forms that satisfy an integrability condition related to the curvature characteristics of the manifold \cite{Leitner}. Moreover, special KY forms and special closed CKY forms which satisfy special types of integrability conditions constitute graded Lie algebra structures on all types of manifolds when they exist.

Hidden symmetries in the presence of torsion has also been studied recently by extending the defining equations of KY and CKY forms to include the torsion of the background manifold \cite{HouriKubiznakWarnickYasui, KubiznakKunduriYasui}. Torsionful geometries are important in backgrounds for supergravity and string theories in various dimensions \cite{FriedrichIvanov}. The geometric description of supergravity theories can be characterized in terms of generalized geometry and the connections in generalized geometry naturally include skew-symmetric torsion 3-form of the background \cite{Hitchin, Gualtieri, CoimbraConstableWaldram, CoimbraConstableWaldram2}. So, hidden symmetries in the presence of torsion also play an important role in generalized geometry. Defining equations for hidden symmetries in generalized geometry are constructed in terms of skew-symmetric torsion in \cite{AcikErtemKelekci}. However, the Lie algebra structure of hidden symmetries in the presence of torsion is not investigated in the literature and still an open problem. This Lie algebra structure can lead to the construction of symmetry superalgebras in torsionful geometries and generalized geometry.

In this paper, we consider CKY forms in the presence of torsion. After finding the integrability conditions of torsionful CKY equation we propose a graded Lie bracket to satisfy a Lie algebra structure for CKY forms with torsion in some special cases. We find that if the skew-symmetric torsion is closed and parallel, which is compatible with the examples in supergravity and string theories, a special subset of CKY forms can satisfy a graded Lie algebra structure on constant curvature or Einstein manifolds. This special subset of CKY forms correspond to CKY forms that satisfy a parallelity condition with respect to torsion operator. Einstein manifolds with parallel skew-symmetric torsion can also be obtained as deformations of Einstein-Sasaki manifolds. We also consider the graded Lie algebra structure of generalized hidden symmetries in generalized geometry which has similar properties with torsionful geometries.

The paper is organized as follows. In Section 2, we consider the manifolds with skew-symmetric torsion and define the connections with skew-symmetric torsion. Section 3 includes the definition of CKY forms with torsion and the integrability conditions for torsionful CKY equation. In Section 4, we define a graded Lie bracket for the Lie algebra structure of torsionful CKY forms and find the conditions to satisfy the graded Lie algebra structure. Section 5 deals with the case of generalized geometry and the graded Lie algebra structure of generalized hidden symmetries. Section 6 concludes the paper.

\section{Connection with Skew-Symmetric Torsion}

We consider a pseudo-Riemannian manifold $(M,g)$ with totally skew-symmetric torsion 3-form $H$. A connection with skew-symmetric torsion $\nabla^H$ can be defined in terms of the Levi-Civita connection $\nabla$ as follows
\begin{equation}
\nabla^H_XY=\nabla_XY+\frac{1}{2}\widetilde{i_Xi_YH}
\end{equation}
where $X$,$Y$ are arbitrary vector fields, $i_X$ denotes the contraction with respect to the vector field $X$ and $\ \widetilde{\,}$ denotes the metric dual operation. In the presence of torsion, for the orthonormal vector field basis $\{X_a\}$ and 1-form basis $\{e^a\}$, we can choose local coordinates that satisfy the following equalities
\begin{equation}
\nabla^H_{X_a}e^b=0
\end{equation}
\begin{equation}
[X_a, X_b]=-T(X_a,X_b)=(i_{X_a}i_{X_b}T^c)X_c
\end{equation}
where $[\,,\,]$ denotes the Lie bracket of vector fields, $T(X,Y)=\nabla^H_XY-\nabla^H_YX-[X,Y]$ is the torsion operator and $T^a=\frac{1}{2}i_{X^a}H$ are torsion 2-forms.

When the connection $\nabla^H$ acts on a differential form $\alpha$, it can be written as
\begin{equation}
\nabla^H_X\alpha=\nabla_X\alpha- \frac{1}{2}i_XH\underset{1}{\wedge}\alpha
\end{equation}
where $i_XH\underset{1}{\wedge}\alpha=i_{X^a}i_XH\wedge i_{X^a}\alpha$. In general, we use the notation
\begin{equation}
\alpha\underset{k}{\wedge}\beta=i_{X_{a_k}}...i_{X_{a_1}}\alpha\wedge i_{X^{a_k}}...i_{X^{a_1}}\beta.
\end{equation}
We can also define operators $d^H$ and $\delta^H$ from the torsionful connection $\nabla^H$ in terms of exterior derivative $d=e^a\wedge\nabla_{X_a}$ and coderivative $\delta=-i_{X^a}\nabla_{X_a}$ operators as follows
\begin{eqnarray}
d^H\alpha&=&e^a\wedge\nabla^H_{X_a}\alpha\nonumber\\
&=&e^a\wedge\nabla_{X_a} \alpha-\frac{1}{2}e^a\wedge(i_{X_a}H\underset{1}{\wedge}\alpha)\nonumber\\
&=&d\alpha+H\underset{1}\wedge\alpha\nonumber\\
\end{eqnarray}
and
\begin{eqnarray}
\delta^H\alpha&=&-i_{X^a}\nabla^H_{X_a}\alpha\nonumber\\
&=&-i_{X^a}\nabla_{X_a}\alpha+\frac{1}{2}i_{X^a}(i_{X_a}H\underset{1}\wedge\alpha)\nonumber\\
&=&\delta\alpha+\frac{1}{2}H\underset{2}\wedge\alpha
\end{eqnarray}
where we have used the relation $e^a\wedge i_{X_a}\alpha=p\alpha$ for a $p$-form $\alpha$. Moreover, we write the curvature operator in the presence of torsion for arbitrary vector fields $X,Y$ as follows
\begin{equation}
R^H(X,Y)=\nabla^H_X\nabla^H_Y-\nabla^H_Y\nabla^H_X-\nabla^H_{[X,Y]}.
\end{equation}
If we consider the basis vector fields $\{X_a\}$, then we can write the curvature operator with torsion acting on a $p$-form $\alpha$ in terms of the torsionless curvature operator $R(X_a, X_b)$ and torsion 3-form $H$ as in the following form
\begin{eqnarray}
R^H(X_a, X_b)\alpha&=&\nabla^H_{X_a}\nabla^H_{X_b}\alpha-\nabla^H_{X_b}\nabla^H_{X_a}\alpha-\nabla^H_{[{X_a},{X_b}]}\alpha\nonumber\\
&=&R(X_a,X_b)\alpha -\frac{1}{2}\left(i_{X_b}\nabla_{X_a}H\underset{1}{\wedge}\alpha-i_{X_a}\nabla_{X_b}H\underset{1}{\wedge}\alpha-i_{[X_a,X_b]}H\underset{1}{\wedge}\alpha\right)\nonumber\\
&&+\frac{1}{4}\left(i_{X_a}H\underset{1}{\wedge}(i_{X_b}H\underset{1}{\wedge}\alpha)-i_{X_b}H\underset{1}{\wedge}(i_{X_a}H\underset{1}{\wedge}\alpha)\right)
\end{eqnarray}
where we have used the identity $[i_X,\nabla_Y]=i_{\nabla_XY}$ and its right hand side vanishes for basis vector fields in normal coordinates as a result of (2).

\section{CKY Forms with Torsion and Integrability Conditions}

On an $n$-dimensional manifold $M$, metric is preserved up to a conformal factor along the flows of conformal Killing vector fields. The antisymmetric generalizations of conformal Killing vector fields to higher degree forms are called conformal Killing-Yano (CKY) forms. A $p$-form $\alpha$ is a CKY $p$-form, if it satisfies the following equation with respect to the Levi-Civita connection $\nabla$ for any vector field $X$
\begin{equation}
\nabla_X\alpha=\frac{1}{p+1}i_Xd\alpha-\frac{1}{n-p+1}\widetilde{X}\wedge\delta\alpha.
\end{equation}
On a manifold with skew-symmetric torsion 3-form $H$, we can generalize the above definition to the torsionful CKY forms in terms of the connection with torsion defined in (4). So, $\alpha$ is a torsionful CKY $p$-form if it satisfies the following equation
\begin{equation}
\nabla^H_X\alpha=\frac{1}{p+1}i_Xd^H\alpha-\frac{1}{n-p+1}\widetilde{X}\wedge\delta^H\alpha.
\end{equation}
By taking the second derivatives of (11), we can find the integrability conditions for torsionful CKY forms. Then, for basis vector fields $\{X_a\}$ satisfying (2) and (3), we can write following equalities from (11)
\begin{eqnarray}
\nabla^H_{X_a}\nabla^H_{X_b}\alpha&=&\frac{1}{p+1}i_{X_b}\nabla^H_{X_a}d^H\alpha-\frac{1}{n-p+1}e_b\wedge\nabla^H_{X_a}\delta^H\alpha\\
\nabla^H_{X_b}\nabla^H_{X_a}\alpha&=&\frac{1}{p+1}i_{X_a}\nabla^H_{X_b}d^H\alpha-\frac{1}{n-p+1}e_a\wedge\nabla^H_{X_b}\delta^H\alpha\\
\nabla^H_{[X_a,X_b]}\alpha&=&-\frac{1}{p+1}i_{T(X_a,X_b)}d^H\alpha+\frac{1}{n-p+1}\widetilde{T(X_a,X_b)}\wedge\delta^H\alpha
\end{eqnarray}
and by combining the equalities (12)-(14) and using the definition of the curvature operator with torsion in (9), we can write
\begin{eqnarray}
R^ H(X_a, X_b)\alpha&=&\frac{1}{p+1}\left(i_{X_b}\nabla^H_{X_a}-i_{X_a}\nabla^H_{X_b}+i_{T(X_a,X_b)}\right)d^H\alpha\nonumber\\
&&-\frac{1}{n-p+1}\left(e_b\wedge\nabla^H_{X_a}-e_a\wedge\nabla^H_{X_b}+\widetilde{T(X_a,X_b)\wedge}\right)\delta^H\alpha.
\end{eqnarray}
By taking the wedge product of (15) from the left by $e^a$, we find
\begin{eqnarray}
e^a\wedge R^H(X_a,X_b)\alpha&=&-\frac{p}{p+1}\nabla^H_{X_b}d^H\alpha-\frac{1}{p+1}i_{X_b}(d^H)^2\alpha+\frac{1}{n-p+1}e_b\wedge d^H\delta^H\alpha\nonumber\\
&&-e^a\wedge\left(-\frac{p}{p+1}i_{T(X_a,X_b)}d^H\alpha+\frac{1}{n-p+1}\widetilde{T(X_a,X_b)}\wedge\delta^H\alpha\right)
\end{eqnarray}
where we have used the identities $e^a\wedge i_{X_a}\alpha=p\alpha$ and $e^a\wedge\nabla^H_{X_a}=d^H$. Moreover, by taking the interior product of (15) with $i_{X^a}$, we find
\begin{eqnarray}
i_{X^a}R^H(X_a,X_b)\alpha&=&\frac{1}{p+1}i_{X_b}\delta^H d^H\alpha+\frac{n-p}{n-p+1}\nabla^H_{X_b}\delta^H\alpha-\frac{1}{n-p+1}e_b\wedge(\delta^H)^2\alpha\nonumber\\
&&+\frac{1}{p+1}i_{X^a}i_{T(X_a,X_b)}d^H\alpha-\frac{1}{n-p+1}i_{X^a}(\widetilde{T(X_a,X_b)}\wedge\delta^H\alpha).
\end{eqnarray}
Then, we obtain from (16)
\begin{eqnarray}
\nabla^H_{X_b}d^H\alpha&=&\frac{p+1}{p(n-p+1)}e_b\wedge d^H\delta^H\alpha-\frac{1}{p}(d^H)^2\alpha\nonumber\\
&&-\frac{p+1}{p}e^a\wedge \widehat{R}(X_a,X_b)\alpha
\end{eqnarray}
and from (17)
\begin{eqnarray}
\nabla^H_{X_b}\delta^H\alpha&=&-\frac{n-p+1}{(p+1)(n-p)}i_{X_b}\delta^H d^H\alpha+\frac{1}{n-p}e_b\wedge(\delta^H)^2\alpha\nonumber\\
&&+\frac{n-p+1}{n-p}i_{X^a}\widehat{R}(X_a,X_b)\alpha
\end{eqnarray}
where we have defined the quantity
\begin{equation}
\widehat{R}(X_a,X_b)\alpha=R^H(X_a,X_b)\alpha+\nabla^H_{T(X_a,X_b)}\alpha.
\end{equation}
By taking the interior product with $-i_{X^b}$ of (18) and wedge product with $e^b$ of (19) and combining them we arrive at
\begin{equation}
\frac{p}{p+1}\delta^Hd^H\alpha+\frac{n-p}{n-p+1}d^H\delta^H\alpha=-e^a\wedge i_{X^b}\widehat{R}(X_a,X_b)\alpha.
\end{equation}
The equalities (18), (19) and (21) are the main integrability conditions for the torsionful CKY equation (11). We can also write these conditions more explicitly in terms of curvature characteristics and torsion 2-forms. The action of curvature operator on forms can be written as \cite{BennTucker}
\begin{equation}
R^H(X_a,X_b)\alpha=[\mathcal{R}_{X_aX_b},\alpha]_{Cl}
\end{equation}
where $[,\,,]_{Cl}$ denotes the Clifford bracket and $\mathcal{R}_{X_aX_b}$ is defined in terms of the curvature 2-forms $R_{ab}$ as
\begin{equation}
\mathcal{R}_{X_aX_b}=-\frac{1}{4}(i_{X_a}i_{X_b}R_{cd})e^{cd}.
\end{equation}
By using the contracted Bianchi identity in the presence of torsion \cite{BennTucker}
\begin{equation}
2\left(i_{X_a}i_{X_q}R_{pr}-i_{X_p}i_{X_r}R_{aq}\right)=\underset{p,q,r,a}{\mathcal{S}}i_{X_p}i_{X_q}i_{X_r}DT_a
\end{equation}
we can write
\begin{eqnarray}
i_{X_a}i_{X_b}R_{cd}&=&i_{X_c}i_{X_d}R_{ab}+\frac{1}{2}\underset{c,b,d,a}{\mathcal{S}}i_{X_c}i_{X_b}i_{X_d}DT_a\nonumber\\
&=&i_{X_c}i_{X_d}R_{ab}+\frac{1}{2}\left(i_{X_c}i_{X_b}i_{X_d}DT_a+i_{X_b}i_{X_d}i_{X_a}DT_c+i_{X_d}i_{X_a}i_{X_c}DT_b+i_{X_a}i_{X_c}i_{X_b}DT_d\right)
\end{eqnarray}
where $\mathcal{S}$ denotes the symmetric sum over the relevant indices and $D$ denotes the covariant exterior derivative. Then, we have
\begin{eqnarray}
-\frac{1}{4}e^{cd}(i_{X_a}i_{X_b}R_{cd})&=&-\frac{1}{4}e^{cd}(i_{X_c}i_{X_d}R_{ab})\nonumber\\
&&-\frac{1}{8}e^{cd}\left(i_{X_c}i_{X_b}i_{X_d}DT_a+i_{X_b}i_{X_d}i_{X_a}DT_c+i_{X_d}i_{X_a}i_{X_c}DT_b+i_{X_a}i_{X_c}i_{X_b}DT_d\right)\nonumber\\
&=&\frac{1}{2}R_{ab}-\frac{1}{8}\left(2i_{X_b}DT_a-e^c\wedge i_{X_b}i_{X_a}DT_c-2i_{X_a}DT_b+e^d\wedge i_{X_a}i_{X_b}DT_d\right)\nonumber\\
&=&\frac{1}{2}R_{ab}+\frac{1}{4}\overline{T}_{ab}
\end{eqnarray}
where we have defined
\begin{equation}
\overline{T}_{ab}:=i_{X_a}DT_b-i_{X_b}DT_a-e^c\wedge i_{X_a}i_{X_b}DT_c.
\end{equation}
Then, we can write (22) as
\begin{eqnarray}
R^H(X_a,X_b)\alpha&=&\frac{1}{2}[R_{ab}+\frac{1}{2}\overline{T}_{ab},\alpha]_{Cl}\nonumber\\
&=&-R_{ab}\underset{1}{\wedge}\alpha-\frac{1}{2}\overline{T}_{ab}\underset{1}{\wedge}\alpha.
\end{eqnarray}
In the last line, we have used the following identity for a 2-form $\beta$ and a $p$-form $\alpha$
\begin{equation}
[\beta,\alpha]_{Cl}=-2\beta\underset{1}{\wedge}\alpha.
\end{equation}
Then, we can write the following equalities for the curvature operator from (16) and (17)
\begin{equation}
e^a\wedge R^{H}(X_a,X_b)\alpha=(-R_{cb}-i_{X_b}DT_c+2e^a\wedge i_{X_c}i_{X_b}DT_a)\wedge i_{X^c}\alpha
\end{equation}
\begin{equation}
i_{X^a}R^H(X_a,X_b)\alpha=P_b\underset{1}{\wedge}\alpha-R_{ab}\underset{1}{\wedge}i_{X^a}\alpha+i_{X_b}i_{X_a}DT_a\underset{1}{\wedge}\alpha-\frac{1}{2}\overline{T}_{ab}\underset{1}{\wedge}i_{X^a}\alpha.
\end{equation}
where $P_b$ is the Ricci 1-form given by $P_b := i_{X^a} R_{ab}$. So, we can write the integrability conditions (18), (19) and (21) in an explicit form as follows
\begin{eqnarray}
\nabla^H_{X_b}d^H\alpha&=&\frac{p+1}{p(n-p+1)}e_b\wedge d^H\delta^H\alpha-\frac{1}{p}i_{X_b}(d^H)^2\alpha\nonumber\\
&&+\frac{p+1}{p}\left(-R_{cb}-i_{X_b}DT_c+2e^a\wedge i_{X_c}i_{X_b}DT_a\right)\wedge i_{X^c}\alpha\nonumber\\
&&+\frac{p+1}{p}i_{X_b}T^c\wedge \nabla^H_{X_c}\alpha
\end{eqnarray}
\begin{eqnarray}
\nabla^H_{X_b}\delta^H\alpha&=&-\frac{n-p+1}{(p+1)(n-p)}i_{X_b}\delta^Hd^H\alpha+\frac{1}{n-p}e_b\wedge(\delta^H)^2\alpha\nonumber\\
&&+\frac{n-p+1}{n-p}\left(P_b\underset{1}{\wedge}\alpha-R_{ab}\underset{1}{\wedge}i_{X^a}\alpha+i_{X^b}i_{X^a}DT_a\underset{1}{\wedge}\alpha-\frac{1}{2}\overline{T}_{ab}\underset{1}{\wedge}i_{X^a}\alpha\right)\nonumber\\
&&-\frac{n-p+1}{n-p}(i_{X_a}i_{X_b}T^c)i_{X^a}\nabla^H_{X_c}\alpha
\end{eqnarray}
\begin{eqnarray}
\frac{p}{p+1}\delta^Hd^H\alpha+\frac{n-p}{n-p+1}d^H\delta^H\alpha&=&P_c\wedge i_{X^c}\alpha+R_{ca}\wedge i_{X^c}i_{X^a}\alpha+3DT_a\underset{1}{\wedge}i_{X^a}\alpha\nonumber\\
&&-\frac{1}{2}e^b\wedge(\overline{T}_{ab}\underset{1}{\wedge}i_{X^a}\alpha)+2T^c\underset{1}{\wedge}\nabla^H_{X_c}\alpha
\end{eqnarray}
where we have used the equality (3).

\section{CKY Bracket and Algebra Structure}

The conditions for satisfying a graded Lie algebra structure of CKY forms were investigated in \cite{Ertem3} . It was proved that for a CKY $p$-form $\alpha_1$ and a CKY $q$-form $\alpha_2$ the following CKY bracket
\begin{eqnarray}
[\alpha_1,\alpha_2]_{CKY}&=&\frac{1}{q+1}i_{X^a}\alpha_1\wedge i_{X_a}d\alpha_2+\frac{(-1)^p}{p+1}i_{X^a}d\alpha_1\wedge i_{X_a}\alpha_2\nonumber\\
&&+\frac{(-1)^p}{n-q+1}\alpha_1\wedge\delta\alpha_2+\frac{1}{n-p+1}\delta\alpha_1\wedge\alpha_2
\end{eqnarray}
is a graded Lie bracket on constant curvature manifolds and the same is true for normal CKY forms on Einstein manifolds. Note that $[\alpha_1,\alpha_2]_{CKY}$ is a $(p+q-1)$-form. Normal CKY forms are defined as the subset of CKY forms satisfying the following integrability conditions
\begin{eqnarray}
\nabla_{X_a}d\alpha&=&\frac{p+1}{p(n-p+1)}e_a\wedge d\delta\alpha+2(p+1)K_a\wedge\alpha\\
\nabla_{X_a}\delta\alpha&=&-\frac{n-p+1}{(p+1)(n-p)}i_{X_b}\delta d\alpha-2(n-p+1)i_{X^b}K_a\wedge i_{X_b}\alpha\\
\frac{p}{p+1}\delta d\alpha+\frac{n-p}{n-p+1}d\delta\alpha&=&-2(n-p)K_a\wedge i_{X^a}\alpha
\end{eqnarray}
where $K_a$ is defined as
\begin{equation}
K_a=\frac{1}{n-2}\left(\frac{\mathcal{R}}{2(n-1)}e_a-P_a\right).
\end{equation}

We can also determine the conditions for the torsionful CKY forms satisfying (11) to constitute a graded Lie algebra structure with respect to a modified CKY bracket. We propose the following $H$-modified graded Lie bracket for torsionful CKY forms
\begin{eqnarray}
[\alpha_1,\alpha_2]_{HCKY}&=&\frac{1}{q+1}i_{X^a}\alpha_1\wedge i_{X_a}d^H\alpha_2+\frac{(-1)^p}{p+1}i_{X^a}d^H\alpha_1\wedge i_{X_a}\alpha_2\nonumber\\
&&+\frac{(-1)^p}{n-q+1}\alpha_1\wedge\delta^H\alpha_2+\frac{1}{n-p+1}\delta^H\alpha_1\wedge\alpha_2.
\end{eqnarray}
To obtain a graded Lie algebra structure for torsionful CKY forms, we need to check that under which conditions the following equality is satisfied for any torsionful CKY $p$-form $\alpha_1$ and torsionful CKY $q$-form $\alpha_2$
\begin{equation}
\nabla^H_{X_a}[\alpha_1,\alpha_2]_{HCKY}=\frac{1}{p+q}i_{X_a}d^H[\alpha_1,\alpha_2]_{HCKY}-\frac{1}{n-p+q+2}e_a\wedge\delta^H[\alpha_1,\alpha_2]_{HCKY}.
\end{equation}
For the left hand side of (41), we can find the following equality by using the torsionful CKY equation (11) and the relevant integrability conditions (18), (19) and (21)
\begin{eqnarray}
\nabla^H_{X_a}[\alpha_1,\alpha_2]_{HCKY}&=&-\frac{1}{(p+1)(q+1)}i_{X_a}(d^H\alpha_1\underset{1}{\wedge}d^H\alpha_2)\nonumber\\
&&+\frac{1}{q(n-q+1)}i_{X_a}\alpha_1\wedge d^H\delta^H\alpha_2+\frac{(-1)^p}{p(n-p+1)}d^H\delta^H\alpha_1\wedge i_{X_a}\alpha_2\nonumber\\
&&-\frac{1}{(p+1)(n-p)}i_{X_a}\delta^Hd^H\alpha_1\wedge\alpha_2-\frac{(-1)^p}{(q+1)(n-q)}\alpha_1\wedge i_{X_a}\delta^Hd^H\alpha_2\nonumber\\
&&+e_a\wedge\bigg(\frac{1}{(n-p+1)(q+1)}\delta^H\alpha_1\underset{1}{\wedge}d^H\alpha_2+\frac{1}{(p+1)(n-q+1)}d^H\alpha_1\underset{1}{\wedge}\delta^H\alpha_2\nonumber\\
&&+\frac{(-1)^p}{q(n-q+1)}\alpha_1\underset{1}{\wedge}d^H\delta^H\alpha_2-\frac{(-1)^p}{p(n-p+1)}d^H\delta^H\alpha_1\underset{1}{\wedge}\alpha_2\nonumber\\
&&+\frac{1}{(n-q)(n-q+1)}\alpha_1\wedge(\delta^H)^2\alpha_2+\frac{1}{(n-p)(n-p+1)}(\delta^H)^2\alpha_1\wedge\alpha_2\bigg)\nonumber\\
&&-\frac{1}{q(q+1)}\alpha_1\underset{1}{\wedge}i_{X_a}(d^H)^2\alpha_2-\frac{(-1)^p}{p(p+1)}i_{X_a}(d^H)^2\alpha_1\underset{1}{\wedge}\alpha_2\nonumber\\
&&-\frac{1}{q}i_{X^b}\alpha_1\wedge \widehat{R}(X_b,X_a)\alpha_2-\frac{(-1)^p}{p}\widehat{R}(X_b,X_a)\alpha_1\wedge i_{X^b}\alpha_2\nonumber\\
&&(-1)^p e^b\wedge\bigg(-\frac{1}{q}\alpha_1\underset{1}{\wedge}\widehat{R}(X_b,X_a)\alpha_2)+\frac{1}{p}\widehat{R}(X_b,X_a)\alpha_1\underset{1}{\wedge}\alpha_2\bigg)\nonumber\\
&&+\frac{(-1)^p}{n-q}\alpha_1\wedge i_{X^b}\widehat{R}(X_b,X_a)\alpha_2+\frac{1}{n-p}i_{X^b}\widehat{R}(X_b,X_a)\alpha_1\wedge \alpha_2.
\end{eqnarray}
The first term on the right hand side of (41) can be found by considering the wedge product of (42) with $e^a$ and using the integrability condition (21). Then, we have
\begin{eqnarray}
\frac{1}{p+q}i_{X_a}d^H[\alpha_1,\alpha_2]_{HCKY}&=&-\frac{1}{(p+1)(q+1)}i_{X_a}(d^H\alpha_1\underset{1}{\wedge}d^H\alpha_2)\nonumber\\
&&+\frac{1}{q(n-q+1)}i_{X_a}\alpha_1\wedge d^H\delta^H\alpha_2+\frac{(-1)^p}{p(n-p+1)}d^H\delta^H\alpha_1\wedge i_{X_a}\alpha_2\nonumber\\
&&-\frac{1}{(p+1)(n-p)}i_{X_a}\delta^Hd^H\alpha_1\wedge\alpha_2-\frac{(-1)^p}{(q+1)(n-q)}\alpha_1\wedge i_{X_a}\delta^Hd^H\alpha_2\nonumber\\
&&\frac{1}{p+q}\bigg(-\frac{(-1)^pp}{q(n-q)}\alpha_1\wedge i_{X_a}(e^c\wedge i_{X^b}\widehat{R}(X_c,X_b)\alpha_2)\nonumber\\
&&-\frac{q}{p(n-p)}i_{X_a}(e^c\wedge i_{X^b}\widehat{R}(X_c,X_b)\alpha_1)\wedge\alpha_2\bigg)\nonumber\\
&&+\frac{(-1)^p}{p+q}e^c\wedge\bigg(\frac{1}{(n-p)} i_{X^b}\widehat{R}(X_c,X_b)\alpha_1\wedge i_{X_a}\alpha_2-\frac{1}{(n-q)}i_{X_a}\alpha_1\wedge i_{X^b}\widehat{R}(X_c,X_b)\alpha_2\bigg)\nonumber\\
&&-\frac{(-1)^p}{q(p+q)}i_{X_a}\bigg(\alpha_1\underset{1}{\wedge}(d^H)^2\alpha_2\bigg)+\frac{1}{p(p+q)}i_{X_a}\bigg((d^H)^2\alpha_1\underset{1}{\wedge}\alpha_2\bigg)\nonumber\\
&&\frac{1}{q(p+q)}\bigg(-i_{X^b}\alpha_1\wedge\widehat{R}(X_b,X_a)\alpha_2+e^b\wedge i_{X_a}(i_{X_c}\alpha_1\wedge\widehat{R}(X_c,X_b)\alpha_2)\bigg)\nonumber\\
&&\frac{(-1)^p}{p(p+q)}\bigg(-\widehat{R}(X_b,X_a)\alpha_1\wedge i_{X^b}\alpha_2+e^b\wedge i_{X_a}(\widehat{R}(X_c,X_b)\alpha_1\wedge i_{X^c}\alpha_2)\bigg)\nonumber\\
&&-\frac{(-1)^p}{q(p+q)}e^c\wedge\bigg(2\alpha_1\underset{1}{\wedge}\widehat{R}(X_c,X_a)\alpha_2-e^{b}\wedge i_{X_a}(\alpha_1\underset{1}{\wedge}\widehat{R}(X_c,X_b)\alpha_2)\bigg)\nonumber\\
&&+\frac{(-1)^p}{p(p+q)}e^c\wedge\bigg(2\widehat{R}(X_c,X_a)\alpha_1\underset{1}{\wedge}\alpha_2-e^{b}\wedge i_{X_a}(\widehat{R}(X_c,X_b)\alpha_1\underset{1}{\wedge}\alpha_2)\bigg)\nonumber\\
&&+\frac{(-1)^p}{(p+q)(n-q)}\bigg(\alpha_1\wedge i_{X^b}\widehat{R}(X_b,X_a)\alpha_2-e^b\wedge i_{X_a}(\alpha_1\wedge i_{X^c}\widehat{R}(X_c,X_b)\alpha_2)\bigg)\nonumber\\
&&+\frac{1}{(p+q)(n-p)}\bigg(i_{X^b}\widehat{R}(X_b,X_a)\alpha_1\wedge\alpha_2-e^b\wedge i_{X_a}(i_{X^c}\widehat{R}(X_c,X_b)\alpha_1\wedge\alpha_2)\bigg)
\end{eqnarray}
The second term on the right hand side of (41) can also be written by considering the contraction of (42) with $i_{X^a}$ as follows
\begin{eqnarray}
-\frac{1}{n-p-q+2}e^a\wedge\delta^H[\alpha_1,\alpha_2]_{HCKY}&=&\frac{1}{(n-p+1)(q+1)}e^a\wedge(\delta^H\alpha_1\underset{1}{\wedge}d^H\alpha_2)\nonumber\\
&&+\frac{1}{(n-q+1)(p+1)}e^a\wedge(d^H\alpha_1\underset{1}{\wedge}\delta^H\alpha_2)\nonumber\\
&&+\frac{(-1)^p}{q(n-q+1)}e^a\wedge(\alpha_1\underset{1}{\wedge}d^H\delta^H\alpha_2)-\frac{(-1)^p}{p(n-p+1)}e^a\wedge(d^H\delta^H\alpha_1\underset{1}{\wedge}\alpha_2)\nonumber\\
&&+\frac{1}{q(q+1)(n-p-q+2)}e^a\wedge(\alpha_1\underset{2}{\wedge}(d^H)^2\alpha_2)\nonumber\\
&&+\frac{1}{p(p+1)(n-p-q+2)}e^a\wedge((d^H)^2\alpha_1\underset{2}{\wedge}\alpha_2)\nonumber\\
&&+\frac{1}{(n-q)(n-q+1)}e^a\wedge\alpha_1\wedge(\delta^H)^2\alpha_2\nonumber\\
&&+\frac{1}{(n-p)(n-p+1)}e^a\wedge(\delta^H)^2\alpha_1\wedge\alpha_2\nonumber\\
&&+\frac{(-1)^p}{q(n-q)(n-p-q+2)}e^a\wedge\left(\alpha_1\underset{1}{\wedge}(e^b\wedge i_{X^c}\widehat{R}(X_b,X_c)\alpha_2)\right)\nonumber\\
&&-\frac{(-1)^p}{p(n-p)(n-p-q+2)}e^a\wedge\left((e^b\wedge i_{X^c}\widehat{R}(X_b,X_c)\alpha_1)\underset{1}{\wedge}\alpha_2\right)\nonumber\\
&&+\frac{1}{n-p-q+2}\bigg[ (-1)^{p+1}\frac{p+q-1}{\,p\,}\,\widehat{R}(X_b,X_a)\alpha_{1}\wedge i_{X_b}\alpha_{2} \nonumber\\
&& -\frac{p+q-1}{\,q\,}\,i_{X_b}\alpha_{1}\wedge \widehat{R}(X_b,X_a)\alpha_{2}+(-1)^{p+1}\frac{p+q-2}{\,q\,}\,e^{b}\wedge\!\big(\alpha_{1} \underset{1}{\wedge} \widehat{R}(X_b,X_a)\alpha_{2}\big) \nonumber\\
&& +\,(-1)^{p}\frac{p+q-2}{\,p\,}\,e^{b}\wedge\!\big(\widehat{R}(X_b,X_a)\alpha_{1}\underset{1}{\wedge} \alpha_{2} \big) + \frac{p+q-1}{\,n-p\,}\,i_{X_b}\widehat{R}(X_b,X_a)\alpha_{1}\wedge \alpha_{2}  \nonumber\\
 && +\,(-1)^{p}\frac{p+q-1}{\,n-q\,}\,\alpha_{1}\wedge i_{X_b}\widehat{R}(X_b,X_a)\alpha_{2}  \bigg]
\end{eqnarray}
Now, we can compare the terms found in (42), (43) and (44) to determine the conditions of satisfying a Lie algebra structure for torsionful CKY forms. One can see that the first nine terms on the right hand side of (42) are equal to the sum of the first five terms on the right hand side of (43) and the first four terms on the right hand side of (44). So, the terms including $d^H$, $\delta^H$, $d^H\delta^H$ and $\delta^H d^H$ are equivalent to each other at the equality (41). Note that $(\delta^H)^2$ terms in (42) and (44) are equal to each other and (43) does not include these type of terms which means that $(\delta^H)^2$ terms in (41) also match. One can check that the following identities can be obtained from (18) and (19) for $(d^H)^2$ and $(\delta^H)^2$
\begin{eqnarray}
(d^H)^2\alpha&=&\frac{1}{2}e^{ab}\wedge\widehat{R}(X_a,X_b)\alpha\\
(\delta^H)^2\alpha&=&\frac{1}{2}i_{X^a}i_{X^b}\widehat{R}(X_a,X_b)\alpha.
\end{eqnarray}
So, $(d^H)^2$ terms in (42), (43) and (44) can be written in terms of the curvature operator defined in (20). Then, only the terms depending on the curvature operator defined in (20) remain to satisfy the equality (41). For the torsionless case, $T^a=0$, all the terms including modified curvature operator reduce to the torsionless curvature operator and the equality (41) is satisfied on constant curvature manifolds for all CKY forms and on Einstein manifolds for normal CKY forms as found in \cite{Ertem3}. From definition (20), the modified curvature operator contains two parts which are $R^H(X_a,X_b)$ and $\nabla^H_{T(X_a,X_b)}$ and  as can be seen from (28) the torsional part of the action of $R^H(X_a,X_b)$ on $\alpha$ only contains covariant exterior derivative of torsion $DT_a$. Since $T^a=\frac{1}{2}i_{X^a}H$, we can write the exterior covariant derivative of torsion 2-form from the following identity \cite{BennTucker}
\begin{eqnarray}
DT_a&=&\frac{1}{2}Di_{X_a}H\nonumber\\
&=&\frac{1}{2}\left(-i_{X_a}dH+\nabla_{X_a}H+i_{X_a}T^b\wedge i_{X_b}H\right)\nonumber\\
&=&\frac{1}{2}\left(-i_{X_a}dH+\nabla_{X_a}H-\frac{1}{2}i_{X_a}H\underset{1}{\wedge}H\right)\nonumber\\
&=&\frac{1}{2}\left(-i_{X_a}dH+\nabla^H_{X_a}H\right).
\end{eqnarray}
If $H$ is a closed 3-form $dH=0$ and is parallel with respect to the connection $\nabla^H$ satisfying $\nabla^H_XH=0$, then we have $DT_a=0$ and the curvature terms $R^H(X_a,X_b)$ reduces to $R(X_a,X_b)$ while $(d^H)^2$ terms vanish from (45). So, in that case the modified curvature terms will only contain the terms including $\nabla^H_{T(X_a,X_b)}$. Moreover, if we restrict the torsionful CKY forms to the subset that satisfy the condition
\begin{equation} \label{Hcondition}
\nabla^H_{T(X_a,X_b)}\alpha=0
\end{equation}
then all the modified curvature terms reduce to torsionless curvature operator and the same conditions in the torsionless case will appear to satisfy the Lie algebra structure. Hence, the torsionful CKY forms satisfying (48) constitute a Lie algebra structure with respect to the bracket $[\,,\,]_{HCKY}$ on constant curvature manifolds with parallel skew-symmetric torsion. Additionally, the normal subset of torsionful CKY forms satisfying (48) constitute a graded Lie algebra structure on Einstein manifolds with parallel skew-symmetric torsion.

If the torsion form $H$ is not closed, as can be seen from Eq. (47) that the parallelity condition $\nabla^H_XH=0$ to obtain an algebra structure for CKY forms will be replaced by the following condition
\begin{equation*}
\nabla^H_{X_a}H=i_{X_a}dH.
\end{equation*}
So, for the torsion forms which are not closed and satisfy this special condition, we can still obtain an algebra structure for torsionful CKY forms. If this condition is not satisfied, then the terms including $DT_a$ will not be zero and the algebra structure for CKY forms cannot be satisfied. On the other hand, if $H$ is exact, then it is automatically closed and satisfying the parallelity condition $\nabla^H_XH=0$ is again enough to satisfy the algebra structure.

Condition \eqref{Hcondition} requires the torsionful CKY form $\alpha$ to be parallel along directions generated by the torsion operator $T(X_a, X_b)$. This amounts to $\alpha$ being invariant under infinitesimal transformations driven by the flux $H$. Though seemingly restrictive, \eqref{Hcondition} arises naturally in many flux-compactification backgrounds:
\begin{itemize}
\item On group manifolds (e.g., NS5-brane WZW models $SU(2) \cong S^3$ or $SU(3)$), $H$ is bi-invariant (e.g., the Cartan 3-form). All left-invariant forms—including standard KY forms—satisfy $\nabla^H_{T(\cdot,\cdot)} \alpha = 0$.
\item In nearly-parallel $G_2$ and nearly-Kähler geometries, the canonical torsion 3-form $\varphi$ is $\nabla^T$-parallel \cite{AgricolaSnri,Friedrich}. Hence, $\varphi$-parallel $p$-forms (e.g., the $G_2$ 3-form or $SU(3)$ torsion classes) obey \eqref{Hcondition} .
\item For supersymmetric heterotic/type-II solutions with NS-NS flux, $dH = 0$ and $\nabla^H H = 0$ imply that spinor-bilinear CKY forms are $\nabla^H$-parallel, satisfying \eqref{Hcondition} ~\cite{PapadopoulosTwistedKY}  . Examples include $\mathrm{AdS}_3 \times S^3 \times S^3 \times S^1$ \cite{DonosGauntlettSparksAdS3}; warped $\mathrm{AdS}_5 \times_w M^5_{\text{Sasaki}}$ \cite{GauntlettMartelliSparksWaldramAdS5}  (deformed to parallel torsion); and specific Strominger-system solutions on Calabi-Yau three-folds (where $\nabla^H$ preserves $\alpha$) \cite{StromingerTorsion}.
\end{itemize}
Thus, \eqref{Hcondition}  defines a physically well-motivated subalgebra of torsionful CKY forms in these key contexts.

It is proved in \cite{AgricolaFerreira} that any Einstein manifold with parallel skew-symmetric torsion $H$ has constant scalar curvature. Moreover, any Einstein-Sasaki manifold has deformations into Einstein manifolds with skew-symmetric torsion. So, the manifolds that have parallel skew-symmetric torsion and satisfy the constant curvature or Einstein condition can be found in the literature. Manifolds with parallel skew-symmetric torsion also play important roles in string theory literature \cite{FriedrichIvanov}. In those backgrounds, the graded Lie algebra structure of torsionful CKY forms satisfying (48) can be realized.

As a concrete example for the implementation of integrability conditions and the algebra structure, consider the round  $S^3$ with left--invariant orthonormal coframe $\{e^1,e^2,e^3\}$ satisfying $de^i=\varepsilon_{ijk}\,e^j\wedge e^k$ (unit radius), and take a constant torsion 3-form $H=h\,e^{123} \, (h \in \mathbb{R})$. Then $dH=0$ and, since $g$ and $H$ are left--invariant, we have $\nabla^{H}H=0$. This setting provides a constant curvature Einstein space (with $\mathrm{Ric}=2g$), so all curvature operators that appear in the integrability identities (cf.\ (18)-(21)) are purely algebraic. Choosing the simplest nontrivial case $p=1$ and $\alpha=e^i$, one has $H\underset{1}{\wedge}e^i=i_{X_i}H=h\,\varepsilon_{ijk}\,e^{jk}$ and $\delta^H e^i=0$, hence $d^H e^i = de^i+H\underset{1}{\wedge}e^i=(1+h)\,\varepsilon_{ijk}\,e^{jk}$. The torsionful CKY equation (11) therefore reduces to $\nabla^{H}_{X} e^i=\tfrac{1}{2}\,i_X d^H e^i$, which is solved by the left--invariant coframe. Evaluating the $H$-modified CKY bracket (40) on $\{e^i\}$ then gives $[e^i,e^j]^{H\!\mathrm{CKY}}=-2(1+h)\,\varepsilon_{ijl}\,e^l$, so the span of $\{e^i\}$ is closed under the bracket (as required by (41)). Thus, on this constant-curvature/Einstein background with $dH=0$ and $\nabla^{H}H=0$, the integrability conditions boil down to a finite set of algebraic checks, and one can read off the resulting symmetry algebra directly.

All statements in Sections III-IV are frame-covariant: upon passing from the String to the Einstein frame $g^{(E)} = e^{2w} g^{(S)}$ (with $H$ unchanged), the CKY\(_H\) equation and the $H$-modified CKY bracket (40) retain the same form with $g^{(S)}$ replaced by $g^{(E)}$. In particular, if $\alpha$ is a torsionful CKY $p$-form for $(g^{(S)},H)$, then $e^{(p+1)w}\alpha$ is a torsionful CKY $p$-form for $(g^{(E)},H)$. Closure of the bracket continues to hold under the same conditions, interpreted in the chosen frame (e.g.\ constant curvature or Einstein, and $\nabla^H H=0$).

\section{Algebra Structure for Generalized CKY Forms}

In generalized geometry, one considers the direct sum of tangent and cotangent bundles $E=TM\oplus T^*M$ of a manifold $M$ which is called the generalized  tangent bundle. Sections of $E$ which is denoted by $\Gamma E$ are called generalized vectors and a generalized vector $\mathcal{X}\in\Gamma E$ can be written in terms of a vector field $X\in \Gamma TM$ and a 1-form $\xi\in\Gamma T^*M$ as
\[
\mathcal{X}=X+\xi.
\]
Similar to the Lie bracket $[\,,\,]$ of vector fields on $\Gamma TM$, we can also define a bracket operation on $E$ called Courant bracket as $[\,,\,]_C:\Gamma E\times\Gamma E\rightarrow \Gamma E$ for two generalized vectors $\mathcal{X}=X+\xi$ and $\mathcal{Y}=Y+\eta$,
\begin{equation}\label{eq7}
[\mathcal{X}, \mathcal{Y}]_C=[X,Y]+\mathcal{L}_X\eta-\mathcal{L}_Y\xi-\frac{1}{2}d\left(i_X\eta-i_Y\xi\right).
\end{equation}
On $E$, one can define a bilinear form $<\,,\,>$ which is written for generalized vectors $\mathcal{X}=X+\xi$ and $\mathcal{Y}=Y+\eta$ as follows
\begin{equation}
<\mathcal{X}, \mathcal{Y}>=\frac{1}{2}\big(i_X\eta+i_Y\xi\big)
\end{equation}
A generalized metric $\mathcal{G}$ with the property $\mathcal{G}^2=I$ can be defined on $E$ by a metric splitting $TM\oplus T^*M=V_+\oplus V_-$ where $V_{\pm}$ correspond to $\pm 1$ eigenspaces of $\mathcal{G}$. The metric $\mathcal{G}$ can be written in terms of the projection of the bilinear form onto $V_{\pm}$ denoted by $<\,,\,>_{\pm}$ as
\begin{equation}
\mathcal{G}(\,,\,)=<\,,\,>_+-<\,,\,>_-.
\end{equation}
One can define an isotropic splitting on $E$, such that $s:\Gamma TM\rightarrow \Gamma E$ and this determines a closed 3-form $H$ on $M$ given by
\begin{equation}\label{eq9}
H(X,Y,Z)=2<[s(X),s(Y)]_C,s(Z)>
\end{equation}
for $X,Y,Z\in \Gamma TM$ and this modifies the Courant bracket to the twisted Courant bracket
\begin{equation}\label{eq10}
[\mathcal{X},\mathcal{Y}]_H=[\mathcal{X},\mathcal{Y}]_C-i_Xi_YH.
\end{equation}
In the presence of a non-zero 3-form field $H$, up to fixing of a divergence operator, a metric-compatible torsion-free generalized connection can be defined in the following form \cite{GarciaFernandez}
\begin{equation}\label{eq19}
\mathbb{D}^{H}_{\mathcal{X}}\mathcal{Y}=\nabla_{\pi(\mathcal{X})}\mathcal{Y}+\epsilon\frac{1}{6}i_{\mathcal{Y}}i_{\pi(\mathcal{X})}H
\end{equation}
where $\mathbb{D}^{H}$ acts on the subbundles $V_{\pm}$ for $\epsilon=\pm 1$, respectively. For $\mathcal{X}=X+\xi$, $\mathcal{Y}=Y+\eta\in\Gamma E$, it can be written explicitly as
\begin{equation}\label{eq20}
\mathbb{D}^{H}_{\mathcal{X}}\mathcal{Y}=\nabla_X\left(\begin{array}{cc}Y\\ \eta\end{array}\right)+\epsilon\frac{1}{6}\left(\begin{array}{cc}\widetilde{i_Yi_X H}\\ i_{\widetilde{\eta}}i_X H\end{array}\right).
\end{equation}
For an orthonormal basis of generalized vectors $\mathcal{X}_A$, one can define the $\mathcal{G}$-duals of $\mathcal{X}_A$ corresponding to the basis for generalized 1-forms $\mathcal{E}_A=\mathcal{G}(\mathcal{X}_A)$ and a generalized $p$-form $\mathcal{A}$ can be written as
\[
\mathcal{A}=a^{A_1A_2...A_p}\mathcal{E}_{A_1}\wedge\mathcal{E}_{A_2}\wedge ... \wedge \mathcal{E}_{A_p}
\]
where $a^{A_1A_2...A_p}$ is a function.
By using the generalized connection, one can write the generalized CKY equation for generalized CKY forms in the following form \cite{AcikErtemKelekci}
\begin{equation}\label{eq105}
\mathbb{D}^H_{\mathcal{X}_A}\mathcal{A}=\frac{1}{p+1}i_{\mathcal{X}_A}d^H\mathcal{A}-\frac{1}{2n-p+1}\mathcal{E}_A\wedge\delta^H\mathcal{A}
\end{equation}
where $d^H=e^a\wedge\nabla^H_{X_a}$ and $\delta^H=-i_{X^a}\nabla^H_{X_a}$. So, the similar analysis as in the previous sections applies to the generalized CKY form case and one can use the similar bracket $[\,,\,]_{HCKY}$ defined in (40) for the graded Lie algebra structure of generalized CKY forms. In that case, the graded Lie bracket is defined in terms of generalized geometry operators as follows
\begin{eqnarray}
[\mathcal{A}_1,\mathcal{A}_2]_{HCKY}&=&\frac{1}{q+1}i_{X^a}\mathcal{A}_1\wedge i_{X_a}d^H\mathcal{A}_2+\frac{(-1)^p}{p+1}i_{X^a}d^H\mathcal{A}_1\wedge i_{X_a}\mathcal{A}_2\nonumber\\
&&+\frac{(-1)^p}{n-q+1}\mathcal{A}_1\wedge\delta^H\mathcal{A}_2+\frac{1}{n-p+1}\delta^H\mathcal{A}_1\wedge\mathcal{A}_2
\end{eqnarray}
where $\mathcal{A}_1$ is a generalized $p$-form and $\mathcal{A}_2$ is a generalized $q$-form. The analysis of graded Lie algebra structure of generalized CKY forms reduces to the analysis of CKY forms in the presence of torsion since the generalized connection defined in (55) can be written in the form of a connection with skew-symmetric torsion. So, the similar analysis will give the result that a subset of generalized CKY forms satisfying a parallelity condition with respect to the torsion operator similar to (48) constitute a graded Lie algebra structure for a closed and parallel 3-form field $H$. Moreover, this bracket reduces to the Schouten-Nijenhuis bracket with $H$ for generalized KY forms and Hodge duals of generalized closed CKY forms which are special subsets of generalized KY forms defined in \cite{AcikErtemKelekci}.

In generalized geometry one works on the bundle $E=TM\oplus T^*M$ with the natural $O(n,n)$ pairing
$\langle X+\xi,\,Y+\eta\rangle=\tfrac12\big(i_X\eta+i_Y\xi\big)$.
A $B$–field acts by the orthogonal transformation $e^{B}(X+\xi)=X+\xi+i_XB$, and the $H$–twisted Courant bracket is
\[
[\,X+\xi,\,Y+\eta\,]_H
=\,[X,Y]+\mathcal{L}_X\eta-\mathcal{L}_Y\xi-\tfrac12\,d\!\big(i_X\eta-i_Y\xi\big)
\;+\;i_Xi_YH .
\]
Generalized metric $\mathcal{G}$ is an $O(n,n)$–orthogonal involution that splits $E=C_+\oplus C_-$ with $C_\pm$ maximal positive/negative; equivalently, it is specified by an ordinary metric $g$ and a 2–form $B$ via
\[
C_+=\{\,X+(g+B)(X,\cdot)\,:\,X\in TM\,\},\qquad
C_-=\{\,X-(g-B)(X,\cdot)\,:\,X\in TM\,\}.
\]
This packages $(g,B)$ into a single geometric object and makes $B$–gauge shifts $B\mapsto B+d\Lambda$ manifest as $O(n,n)$ isometries. The metric connections on $TM$ that are compatible with $\mathcal{G}$ are precisely the \emph{Bismut} connections
$\nabla^{\pm}=\nabla^{g}\pm\tfrac12\,g^{-1}H$.
The operators we use are the differential and codifferential modified by $H$,
$d^H=d+H\underset{1}{\wedge}$ and $\delta^H=\delta+\tfrac12 H\underset{2}{\wedge}$; in the generalized picture, $d^H$ is the de~Rham differential twisted by $H$ (the spinor differential for $E$), so
\[
(d^H)^2=0\quad\Longleftrightarrow\quad dH=0 .
\]
Thus the condition $dH=0$ ensures that the $H$–twisted Courant bracket satisfies the Jacobi identity (a genuine Courant algebroid) and that the derived–bracket manipulations underlying our algebra close.
Likewise, the condition $\nabla^{H}H=0$ is the statement that the torsion is parallel for the Bismut connection, i.e. the generalized metric is preserved by a compatible torsionful connection. In our identities this removes all $\nabla^{H}H$–remainders and reduces the curvature terms to the constant–curvature/Einst\-ein pieces used for the closure of the $H$–modified CKY bracket.
In short, the generalized metric viewpoint packages $(g,B)$ and the $H$–twist into one structure where the integrability requirements we impose ($dH=0$, $\nabla^{H}H=0$) and the resulting algebraic closures appear as natural consequences.

\section{Conclusion}

Besides mathematical relevance, hidden symmetries in the presence of torsion are also important in supergravity and string theories in various dimensions. Extension of the Lie algebra structure of isometries for higher degree generalizations can give way to obtain the full symmetry structure of a background. By using these Lie algebra structures one can construct extended symmetry superalgebra structures by including special types of spinors. In this paper, we have investigated the graded Lie algebra structure of CKY forms in the presence of torsion. A graded Lie bracket called HCKY bracket is proposed which corresponds to a slight modification of the SN bracket and reduces exactly to the torsionless case when torsion is taken as zero. We find that when the skew-symmetric torsion 3-form $H$ is closed and parallel then a special subset of CKY forms constitute a graded Lie algebra structure with respect to HCKY bracket on constant curvature or Einstein manifolds. The special subset consists of CKY forms satisfying a special parallelity condition with respect to torsion operator given in (48). It is also shown that for the torsionless case the condition to satisfy a graded Lie algebra structure reduces to the graded Lie algebra structure of CKY forms which is relevant on constant curvature manifolds or on Einstein manifolds for normal CKY forms. We also discuss the case for generalized geometry which includes a generalized connection that can be written in terms of skew-symmetric torsion. It is found that similar graded Lie algebra structure can also be constructed for generalized CKY forms in generalized geometry.

The results found  in the paper can be used in the construction of extended symmetry superalgebra structures for torsionful manifolds. The even parts of symmetry superalgebra structures correspond to the Lie algebras of hidden symmetries and one can also investigate the torsionful Killing and twistor spinors to obtain the odd parts of the superalgebras. Similar structures in generalized geometry can also be investigated by using the graded Lie algebra structures of generalized hidden symmetries.

\begin{acknowledgments}

This study was supported by Scientific and Technological Research Council of T\"urkiye (T\"UB\.ITAK) under the Grant Number 123F261. The authors thank to T\"UB\.ITAK for their supports. Ü. E. thanks to Uwe Semmelmann and Niklas Rauchenberger for valuable discussions during his visit to Stuttgart University, Institute of Geometry and Topology.

\end{acknowledgments}





\begin{thebibliography}{99}

\bibitem{Yano} K. Yano, Some remarks on tensor fields and curvature, Ann. Math. 55 (1952) 328-347.

\bibitem{TachibanaKashiwada} S. Tachibana, T. Kashiwada, On the integrability of Killing-Yano equations, J. Math. Soc. Japan 21 (1969) 259-265.

\bibitem{Semmelmann} U. Semmelmann, Conformal Killing forms on Riemannian manifolds, Math. Z. 245 (2003) 503.

\bibitem{HughstonPenroseSommersWalker} L. P. Hughston, R. Penrose, P. Sommers and M. Walker, On a quadratic first integral for the charged particle orbits in the charged Kerr solution, Commun. Math. Phys. 27 (1972) 303.

\bibitem{KrtousKubiznakPageFrolov} P. Krtous, D. Kubiznak, D. N. Page, V. P. Frolov, Killing-Yano tensors, rank-2 Killing tensors, and conserved quantities in higher dimensions, J. High Energy Phys. 0702 (2007) 004.

\bibitem{BennKress} I. M. Benn, J. Kress, First-order Dirac symmetry operators, Class. Quantum Grav. 21 (2004) 427.

\bibitem{AcikErtemOnderVercin1} \"{O}. A\c{c}{\i}k, \"{U}. Ertem, M. \"{O}nder, A. Ver\c{c}in, First-order symmetries of the Dirac equation in a curved background: a unified dynamical symmetry condition, Class. Quantum Grav. 26 (2009) 075001.

\bibitem{BennCharlton} I. M. Benn, P. Charlton, Dirac symmetry operators from conformal Killing-Yano tensors, Class. Quantum Grav. 14 (1997) 1037.

\bibitem{CarigliaKrtousKubiznak} M. Cariglia, P. Krtous, D. Kubiznak, Dirac equation in Kerr-NUT-(A)dS spacetimes: Intrinsic characterization of separability in all dimensions, Phys. Rev. D 84 (2011) 024008.

\bibitem{Cariglia} M. Cariglia, Hidden symmetries of dynamics in classical and quantum physics, Rev. Mod. Phys. 86 (2014) 1283.

\bibitem{KastorTraschen} D. Kastor, J. Traschen, Conserved gravitational charges from Yano tensors, J. High Energy Phys. 0408 (2004) 045.

\bibitem{AcikErtemOnderVercin2} \"{O}. A\c{c}{\i}k, \"{U}. Ertem, M. \"{O}nder A. Ver\c{c}in, Basic gravitational currents and Killing-Yano forms, Gen. Relativ. Gravit. 42 (2010) 2543.

\bibitem{ErtemAcik} \"{U}. Ertem, \"{O}. A\c{c}{\i}k, Couplings of gravitational currents with Chern-Simons gravities, Phys. Rev. D 87 (2013) 044052.

\bibitem{Ertem} \"U. Ertem, Symmetry operators of Killing spinors and superalgebras in AdS$_5$, J. Math. Phys. 57 (2016) 042502.

\bibitem{Ertem2} \"U. Ertem, Twistor spinors and extended conformal superalgebras, J. Geom. Phys. 152 (2020) 103654.

\bibitem{AcikErtem} \"O. A\c{c}{\i}k, \"U. Ertem, Generalized symmetry superalgebras, J. Math. Phys. 62 (2021) 052303.

\bibitem{KastorRayTraschen} D. Kastor, S. Ray, J. Traschen, Do Killing-Yano tensors form a Lie algebra?, Class. Quantum Grav. 24 (2007) 3759.

\bibitem{Ertem3} \"{U}. Ertem, Lie algebra of conformal Killing-Yano forms, Class. Quantum Grav. 33 (2016) 125033.

\bibitem{CarigliaKrtousKubiznak2} M. Cariglia, P. Krtous, D. Kubiznak, Commuting symmetry operators of the Dirac equation, Killing-Yano and Schouten-Nijenhuis brackets, Phys. Rev. D 84 (2011) 024004.

\bibitem{Leitner} F. Leitner, Conformal Killing forms with normalisation condition, Rend. Circ. Mat. Palermo, Serie II Suppl. 75 (2005) 279-292.

\bibitem{HouriKubiznakWarnickYasui} T. Houri, D. Kubiznak, C. Warnick, Y. Yasui, Symmetries of the Dirac operator with skew-symmetric torsion, Class. Quantum Grav. 27 (2010) 185019.

\bibitem{KubiznakKunduriYasui} D. Kubiznak, H. K. Kunduri, Y. Yasui, Generalized Killing-Yano equations in D=5 gauged supergravity, Phys. Lett. B 678 (2009) 240-45.

\bibitem{FriedrichIvanov} T. Friedrich, S. Ivanov, Parallel spinors and connections with skew-symmetric torsion in string theory, Asian J. Math. 6 (2002) 303-36.

\bibitem{Hitchin} N. Hitchin, Generalized Calabi-Yau manifolds, Q. J. Math. 54 (2003) 281-308.

\bibitem{Gualtieri} M. Gualtieri, Generalized complex geometry, PhD Thesis, University of Oxford, 2004.

\bibitem{CoimbraConstableWaldram} A. Coimbra, C. Strickland-Constable, D. Waldram, Supergravity as generalised geometry I: type II theories, J. High Energy Phys. 1111 (2011) 091.

\bibitem{CoimbraConstableWaldram2} A. Coimbra, C. Strickland-Constable, D. Waldram, Supergravity as generalised geometry II: $E_{d(d)}\times\mathbb{R}^+$ and M-theory, J. High Energy Phys. 1403 (2014) 019b .

\bibitem{AcikErtemKelekci} \"{O}. A\c{c}{\i}k, \"{U}. Ertem, \"{O}. Kelek\c{c}i, Spinor bilinears and Killing-Yano forms in generalized geometry, arXiv:2411.00443.

\bibitem{BennTucker} I. M. Benn, R. W. Tucker, An Introduction to Spinors and Geometry with Applications in Physics, IOP Publishing, Bristol, 1987.

\bibitem{AgricolaSnri} I. Agricola, The Srní Lectures on non-integrable geometries with torsion, \newblock Arch. Math. (Brno) \textbf{42} (2006), 5-84.

\bibitem{Friedrich}T. Friedrich, $G_2$-manifolds with parallel characteristic torsion, Differential Geom. Appl. {\bf 25} (2007), no.~6, 632--648.

\bibitem{PapadopoulosTwistedKY}
G. Papadopoulos, Twisted form hierarchies, Killing–Yano equations and supersymmetric backgrounds, J. High Energy Phys. 2007 (2020) 025.

\bibitem{DonosGauntlettSparksAdS3}
A. Donos, J. P. Gauntlett, J. Sparks, AdS$_3$ $\times$ S$^3$ $\times$ S$^3$ $\times$ S$^1$ solutions of type IIB string theory, Class. Quantum Grav. 26 (2009) 065009.

\bibitem{GauntlettMartelliSparksWaldramAdS5}
J. P. Gauntlett, D. Martelli, J. Sparks, D. Waldram, Supersymmetric AdS$_5$ solutions of type IIB supergravity, Class. Quantum Grav. 23 (2006) 4693–4718.

\bibitem{StromingerTorsion}
A. Strominger, Superstrings with torsion, Nucl. Phys. B 274 (1986) 253–284.

\bibitem{AgricolaFerreira} I. Agricola, A. C. Ferreira, Einstein manifolds with skew torsion, Quart. J. Math. 65 (2014) 717-41.

\bibitem{GarciaFernandez} M. Garcia-Fernandez, Ricci flow, Killing spinors and T-duality in generalized geometry, Adv. Math. 350 (2019) 1059-1108.

\end{thebibliography}
\end{document}